\documentclass[twocolumn,superscriptaddress,prl,aps]{revtex4-1}
\usepackage{amssymb}
\usepackage{amsmath}
\usepackage{graphicx}
\usepackage{color,xcolor}

\begin{document}

\title{Unitary Scattering Protected by Pseudo-Hermiticity}
\author{L. Jin}
\email{jinliang@nankai.edu.cn}
\affiliation{School of Physics, Nankai University, Tianjin 300071, China}

\begin{abstract}
The Hermitian systems possess unitary scattering; however, the Hermiticity
is unnecessary for a unitary scattering although the scattering under the
influence of non-Hermiticity is mostly non-unitary. Here we prove that the
unitary scattering is protected by certain type of pseudo-Hermiticity and
unaffected by the degree of non-Hermiticity. The energy conservation is
violated in the scattering process and recovers after scattering. The
subsystem of the pseudo-Hermitian scattering center including only the
connection sites is Hermitian. These findings provide fundamental insights
on the unitary scattering, pseudo-Hermiticity, and energy conservation; and
are promising for the light propagation, mesoscopic electron transport, and
quantum interference in the non-Hermitian systems.
\end{abstract}

\maketitle

Pseudo-Hermiticity is important in non-Hermitian physics. The
pseudo-Hermiticity ensures the spectrum of the non-Hermitian system to be
entirely real or partly complex in conjugate pairs \cite{Ali02}. A
non-Hermitian system is pseudo-Hermitian if its Hamiltonian under a unitary
transformation equals to the Hermitian conjugation of the Hamiltonian \cite%
{Jones05}. The parity-time ($\mathcal{PT}$) symmetric non-Hermitian systems
are the mostly investigated pseudo-Hermitian systems \cite%
{Bender98,Dorey01,Muga05,LJin09,Joglekar11,Suchkov16,Chong17,Lai21,Chen20,Lv21,Zi21}%
, which possess non-unitary dynamics even their spectra are entirely real.
The state involving nonorthogonal eigenmodes exhibits non-unitary intensity
oscillation as observed in the coupled optical waveguides \cite%
{Ganainy07,Makris08,CERuter}; nevertheless, the eigenstates are orthogonal
and the time-evolution is unitary under the biorthogonal norm \cite%
{Ali04,Brody13}. Interestingly, the state only involving real-valued
orthogonal eigenmodes in the pseudo-Hermitian systems exhibits an
intensity-preserving dynamics \cite{LJinPRA11}. Otherwise, the intensity
exponentially increases/decreases in the broken $\mathcal{PT}$-symmetric
phase \cite{Kottos10,Zheng10} or polynomially increases at the exceptional
point where the $\mathcal{PT}$-symmetric phase transition occurs \cite%
{PWang16,LGe18,Xue19}. The exceptional point in the $\mathcal{PT}
$-symmetric systems is experimentally realized in optical/acoustic cavity
resonators~\cite{LYang14,CTChanPRX16}, in the single-photon interferometric
quantum simulation~\cite{Xue20}, in the single nitrogen-vacancy center~\cite%
{JDu21} and so on~\cite{Ueda20}.

Non-Hermitian systems provide unprecedented opportunities in
recent decades \cite{Moiseyev}. The rapid developments in non-Hermitian
physics greatly stimulate novel applications in optics, condensed matter
physics, quantum physics, and material science \cite%
{Konotop16,Kivshar,LFeng,Longhi,Ganainy,Alu,LYang,YFChen}; for example, the
exceptional point enhanced optical sensing \cite%
{Wiersig14,Hodaei17,Clerk18,Lai19,Cai20}, robust energy transfer \cite%
{Harris16,Fan17}, lasing \cite{Feng14,Harari18}, and many other intriguing
phenomena including the coherence perfect absorption \cite%
{YDChong10,Longhi10,HCao11,HChen14,TKottos17,Jeffers19}, unidirectional
reflectionless/invisibility \cite%
{ZLin2011,Regensburger12,LFeng2013,Zhu14,JHWu14,Sounas15,Makris20},
absorption \cite{LonghiOL15,Sweeney19}, amplification \cite{CLi17,Fleury18},
and lasing were discovered \cite{Ali09,Ramezani14,LJinPRL}. These reveal the
non-unitary feature and the asymmetric feature of scattering affected by the
non-Hermiticity \cite{Muga17}. In addition, the conservation is an important
topic in non-Hermitian physics \cite{Alexandre17,Xue20,LGe20}. The energy
conservation from the unitary scattering was reported in several
non-Hermitian scattering centers \cite%
{LJinPRA12,LGe12,Ahmed13,Mostafazadeh14,LJinPRA18}. Thus, the Hermiticity is
unnecessary for a unitary scattering; however, the non-unitary scattering
more commonly appears in the non-Hermitian systems because of the lack of
energy conservation \cite%
{Muga04,Cannata07,JonesPRD07,ZnojilPRD08,SRotter13,Schomerus13,Kottos15,LGe15,LJin16,Pagneux17,ZZhao19,Droulias19,Novitsky20,Haque20,Schomerus21,Economou21,Krasnok19}%
. Then, what is essential for a unitary scattering and the energy
conservation in non-Hermitian physics? This is a fundamental and important
problem. Here we thoroughly solve this problem and unveil that the
pseudo-Hermiticity plays a vital role.

In this Letter, we report that the unitary scattering in the non-Hermitian
systems is protected by certain type of pseudo-Hermiticity. Under the
pseudo-Hermiticity protection, the total probability of wave injection
remains unity after scattering; whereas the energy conservation is violated
in the scattering process as affected by the non-Hermiticity. We report that
the unitary scattering and energy conservation are independent of the degree
of non-Hermiticity, but strongly depend on the structure of the scattering
center. We provide novel understandings of pseudo-Hermiticity from the
perspective of scattering; and present fundamental insights on the unitary
and non-unitary scattering. Furthermore, the consequences of symmetry
protections on the scattering matrix under the time-reversal symmetry and
reciprocity generalized for the non-Hermitian systems are presented.

\begin{figure}[tb]
\includegraphics[ bb=0 0 310 335, width=4.8 cm, clip]{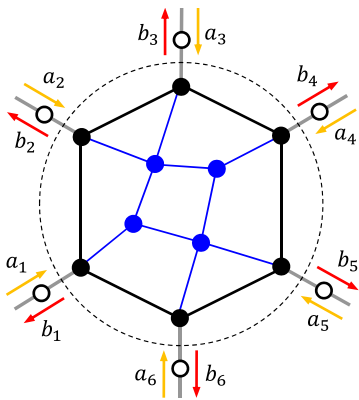}
\caption{Schematic of a general scattering system.
The scattering center is indicated inside the dashed black circle, where the solid circles are the scattering center sites and the solid lines indicate their couplings. The ports in gray are coupled to the connection sites in black, and the other sites disconnected with the ports are the bulk sites in blue. The first sites of the ports are in white. The arrows indicate the incoming and outgoing waves in the ports.}
\label{fig1}
\end{figure}

A lattice model is schematically illustrated in Fig.~\ref{fig1} and
characterizes the discrete systems modeled under the tight-binding
approximation in contrast to the continuous models \cite{Muga19,Muga20};
for example, the coupled resonators/waveguides \cite%
{LFeng2013,Schomerus14}, acoustic crystals \cite{Zhu14,Fleury15,BZhang20},
cold atoms in optical lattice \cite{Gou20}, and electronic circuits \cite%
{Kottos11,Joglekar21}. These experimental platforms are intensively used
for studying the non-Hermitian Hamiltonians. In general, the scattering
center $H_{c}$ has $N$ sites; and all the $L$ ports are connected to
different sites of the scattering center ($L\leq N$).

The properties of the scattering center are fully characterized by the
scattering matrix $S$. Acting the scattering matrix on the incoming wave
amplitudes yields the outgoing wave amplitudes
\begin{equation}
B=SA,  \label{S}
\end{equation}%
where $A=[a_{1},a_{2},\cdots ,a_{L-1},a_{L}]^{T}$ represents the incoming
wave amplitudes of all the $L$ ports before scattering, and $%
B=[b_{1},b_{2},\cdots ,b_{L-1},b_{L}]^{T}$ represents the outgoing wave
amplitudes of all the $L$ ports after scattering. The scattering matrix
element $s_{nm}$ describes the output in the $n$-th port for the wave
injection in the $m$-th port.

The wavefunction at the steady-state is the superposition of the plane waves
with the opposite momenta $k$ propagating in the opposite directions. The
incoming wave is $e^{-ikj}$ and the outgoing wave is $e^{ikj}$ for the
momentum $k$, where the integer $j>0$ indexes the sites of the ports and $%
j=0 $ represents the connection site of the scattering center. To reflect
the properties of the scattering center $H_{c}$, all the ports are chosen
identical and uniform until the scattering center. For the wave injection in
the $m$-th port, the wavefunction in the $m$-th port is $\varphi
_{p,m}(j)=e^{-ikj}+s_{mm}e^{ikj}$ and the wavefunction in the $n$-th port is
$\varphi _{p,n}(j)=s_{nm}e^{ikj}$.

Symmetries are extremely important in physics. The pseudo-Hermitian
scattering center satisfies
\begin{equation}
qH_{c}^{\ast }q^{-1}=H_{c}^{T}.  \label{Hc}
\end{equation}%
In the superscripts, $\ast $ is the complex conjugation operation, $T$ is
the transpose operation, and $q$ is the unitary matrix defined
in the real space representation of the scattering center with $q^{2}=I_{N}$~%
\cite{KawabataPRX19,HZhouPRB19}. The progresses on the symmetry
classification have greatly advanced our knowledge on the symmetric
scattering in non-Hermitian physics \cite{LJinCPL}. In general case, the
pseudo-Hermiticity cannot ensure the unitary scattering; however, the
scattering matrix $S$ is unitary if the unitary matrix $q$ satisfies%
\begin{equation}
q=\left(
\begin{array}{cc}
I_{L} & 0 \\
0 & q_{b}%
\end{array}%
\right) ,  \label{q}
\end{equation}%
where identical matrix $I_{L}$ is for the subsystem including only the $L$
connection sites and the unitary matrix $q_{b}$ is for the subsystem
including only the $N-L$ bulk sites. For all the basis of the scattering
center, the connection sites are exactly mapped to themselves under the
unitary transformation $q$. In Fig.~\ref{fig1}, the black part is Hermitian
and the blue part is pseudo-Hermitian; otherwise, the unitary scattering is
not ensured.

Applying the unitary transformation $q$ to the pseudo-Hermitian scattering
system, the bulk sites (the blue circles in Fig.~\ref{fig1}) and the
couplings between the bulk and connection sites (the blue lines in Fig.~\ref%
{fig1}) are altered, but the subsystem including all the connection sites
(the black circles and lines in Fig.~\ref{fig1}) is unchanged. Consequently,
the wavefunctions of the connection sites and all the ports are invariant
under the unitary transformation $q$. Thus, the scattering matrix for the
scattering center $qH_{c}^{\ast }q^{-1}$ is identical with the scattering
matrix for the scattering center $H_{c}^{\ast }$.

From Eq.~(\ref{S}), we obtain $A^{\ast }=\left( S^{\ast }\right)
^{-1}B^{\ast }$. After applying the complex conjugation operation to the
wavefunctions of the scattering system, the vector $B^{\ast }$ indicates the
incoming wave amplitudes, and the vector $A^{\ast }$ indicates the outgoing
wave amplitudes. Thus, $\left( S^{\ast }\right) ^{-1}$ stands for the
scattering matrix of the scattering center $H_{c}^{\ast }$.

Then, the scattering matrix $\left( S^{\ast }\right) ^{-1}$ is also the
scattering matrix of $qH_{c}^{\ast }q^{-1}$. Notably, the scattering matrix
for $H_{c}^{T}$ is $S^{T}$~\cite{STProof}. From the
pseudo-Hermiticity of $H_{c}$, we obtain the relation $\left( S^{\ast
}\right) ^{-1}=S^{T}$. Therefore, the scattering matrix is unitary
\begin{equation}
SS^{\dagger }=I_{L}.
\end{equation}%
The unity element in the $m$-th row and $m$-th column of $S^{T}S^{\ast }$ is
$\sum_{n=1}^{L}s_{nm}s_{nm}^{\ast }=1$ for the input in the $m$-th port. The
unity diagonal elements of $S^{T}S^{\ast }$ yield the unity total
probability and energy conservation after scattering for the input in any
port. However, the dynamics is non-unitary and the energy conservation is
invalid as affected by the non-Hermiticity in the scattering process.
Notably, the pseudo-Hermiticity-protected unitary scattering is independent
of the degree of non-Hermiticity. Under the pseudo-Hermiticity
protection, the unitary scattering is unaffected by the strengths of
non-Hermitian couplings and the rates of gains/losses; however, these
non-Hermitian elements affect the reflections, the transmissions, and the
dynamics in the scattering process.

The pseudo-Hermitian scattering centers holding unitary scattering have
featured structures. $H_{e}$ denotes the subsystem that only contains the
connection sites; $H_{b}$ denotes the subsystem that only contains the bulk
sites; $H_{eb}$ and $H_{be}$ denote the couplings between the connection
sites and the bulk sites. The scattering center is%
\begin{equation}
H_{c}=\left(
\begin{array}{cc}
H_{e} & H_{eb} \\
H_{be} & H_{b}%
\end{array}%
\right) .
\end{equation}%
The pseudo-Hermiticity $qH_{c}^{\ast }q^{-1}=H_{c}^{T}$ and the block
diagonal $q$ in Eq.~(\ref{q}) yield
\begin{equation}
\left(
\begin{array}{cc}
H_{e}^{\ast } & H_{be}^{\ast }q_{b}^{-1} \\
q_{b}H_{eb}^{\ast } & q_{b}H_{b}^{\ast }q_{b}^{-1}%
\end{array}%
\right) =\left(
\begin{array}{cc}
H_{e}^{T} & H_{eb}^{T} \\
H_{be}^{T} & H_{b}^{T}%
\end{array}%
\right) .  \label{BDEQ}
\end{equation}%
From $q^{2}=I_{N}$, we have $q=q^{-1}$, $q_{b}=q_{b}^{-1}$; and $%
q_{b}(q_{b}^{-1})^{\dagger }=I_{N-L}$. Therefore, $(H_{be}^{\ast
}q_{b}^{-1})^{\dagger }=(H_{eb}^{T})^{\dagger }$ yields $q_{b}(q_{b}^{-1})^{%
\dagger }H_{be}^{T}=q_{b}H_{eb}^{\ast }$.

The unitary scattering requires three constrains%
\begin{equation}
\text{(i) }H_{e}^{\ast }=H_{e}^{T}\text{, (ii) }q_{b}H_{b}^{\ast
}q_{b}^{-1}=H_{b}^{T}\text{, (iii) }q_{b}H_{eb}^{\ast }=H_{be}^{T}\text{.}
\label{Constrain}
\end{equation}%
The pseudo-Hermitian scattering center may have all kinds of
non-Hermitian elements including the gain/loss, the imaginary/complex
coupling, the asymmetric coupling, and etc. These non-Hermitian elements may
simultaneously present in the pseudo-Hermitian scattering center that
possessing the unitary scattering. The constrain (i) requires that the
subsystem $H_{e}$ including only the connection sites is Hermitian. The
constrain (ii) requires that the subsystem $H_{b}$ including only the bulk
sites is pseudo-Hermitian. The constrain (iii) is the requirement on the
couplings $H_{eb}$ and $H_{be}$ between the connection sites and the bulk
sites. Thus, the gain and loss cannot appear at the connection
sites, but can appear on the bulk sites in the balanced pairs; whereas the
non-Hermitian couplings including both the imaginary/complex coupling and
the asymmetric coupling cannot appear between the connection sites, but can
appear among the bulk sites or as the connection couplings in $H_{eb}$ and $%
H_{be}$. Otherwise, the scattering is non-unitary.

The pseudo-Hermiticity-protected two-port scattering centers possess
symmetric transmission and reflection for the wave injections from the
opposite directions ($|t_{L}|^{2}=|t_{R}|^{2}$ and $|r_{L}|^{2}=|r_{R}|^{2}$%
). This is obtained from the unitary scattering $SS^{\dagger }=[1,0;0,1]$
with $S=[r_{L},t_{R};t_{L},r_{R}]$ \cite{SSI}, where $t_{L}$ and $r_{L}$ ($%
t_{R}$ and $r_{R}$) are the transmission and reflection coefficients for the
input in the left (right) port. These explain the unitary and symmetric
scattering in the two-port non-Hermitian scattering centers \cite%
{LGe12,Ahmed13,Mostafazadeh14,LJinPRA18,LJinPRA12}. However, the symmetric
scattering is not promised in the multi-port scattering centers although the
scattering is unitary. In a circulator, the wave injected in the port $1,2,3$
resonantly outgoes from the port $2,3,1$, respectively \cite{Fleury14}. The
scattering is asymmetric when considering the wave input and output in any
two of the three ports.

We have rigorously proved that the unitary scattering and the energy
conservation in the non-Hermitian systems are protected by certain
pseudo-Hermiticity if the unitary matrix $q$ that defined the
pseudo-Hermiticity satisfies Eq.~(\ref{q}). Furthermore, we elaborate a
two-port scattering center to emphasize the importance of the scattering
center configuration and a three-port scattering center to emphasize the
importance of the port configuration for the unitary scattering. In the
schematics, each site stands for a resonator with frequency $\omega _{c}$.
The ports until the scattering center are uniform at the coupling $-J$. The
light propagation in the coupled resonator optical waveguides is govern by
the discrete lattice model \cite{Kippenberg02}, and the dispersion relation
supported by the ports is $E=\omega _{c}-2J\cos k$ \cite{Ali09,Ramezani14}.

\begin{figure}[tb]
\includegraphics[ bb=0 0 400 400, width=8.8 cm, clip]{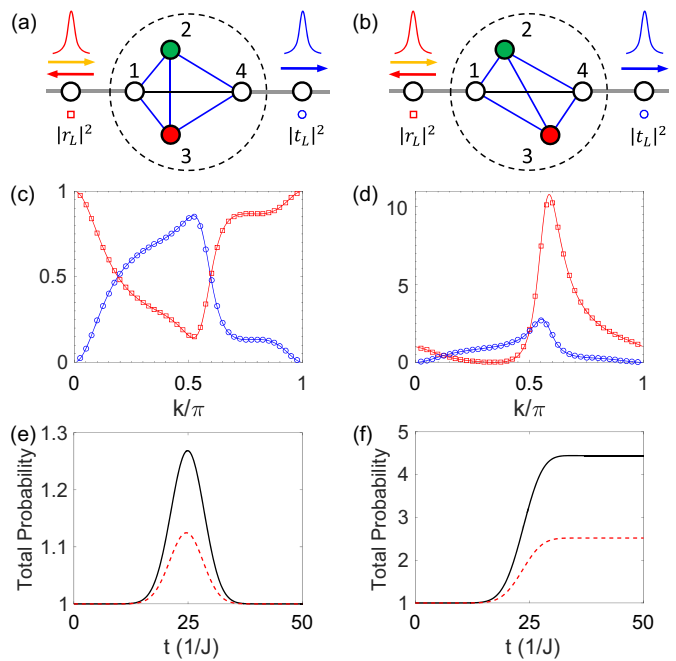}
\caption{Schematics of the two-port pseudo-Hermitian systems with (a) unitary
and (b) non-unitary scattering. The reflection $|r_L|^2$ and transmission $|t_L|^2$ are plotted in (c) and (d)
as indicated by the markers in the ports; and the total probability in the numerical simulations for $k=\pi /2$
are depicted in (e) and (f). In (c) and (d), the curves are analytical results and the markers are numerical simulations.
In (e) and (f), the solid black (dashed red) lines are for the left (right) input. The parameters are $\gamma=J_1=J$, $\kappa_1=\kappa_2=J_2=J/2$.}
\label{fig2}
\end{figure}

The two-port pseudo-Hermitian scattering center in Fig.~\ref{fig2}(a)
includes two resonators with balanced gain and loss ($2$ and $3$) and two
connection resonators ($1$ and $4$). If each connection resonator is equally
coupled to the gain and loss resonators, the equations of motion for the
scattering center are%
\begin{equation}
\begin{array}{l}
i\dot{\psi}_{c,1}=\omega _{c}\psi _{c,1}-J_{1}\psi _{c,2}-J_{1}\psi
_{c,3}-\kappa _{2}\psi _{c,4}-J\psi _{p,1}\left( 1\right) , \\
i\dot{\psi}_{c,2}=\left( \omega _{c}+i\gamma \right) \psi _{c,2}-J_{1}\psi
_{c,1}-\kappa _{1}\psi _{c,3}-J_{2}\psi _{c,4}, \\
i\dot{\psi}_{c,3}=\left( \omega _{c}-i\gamma \right) \psi _{c,3}-J_{1}\psi
_{c,1}-\kappa _{1}\psi _{c,2}-J_{2}\psi _{c,4}, \\
i\dot{\psi}_{c,4}=\omega _{c}\psi _{c,4}-J_{2}\psi _{c,2}-J_{2}\psi
_{c,3}-\kappa _{2}\psi _{c,1}-J\psi _{p,2}\left( 1\right) ,%
\end{array}%
\end{equation}%
where $\psi _{c,j}$ is the wavefunction for the resonator $j$ of the
scattering center and $\psi _{p,j}\left( 1\right) $ is the wavefunction for
the first resonator of the port $j$. The transmission and reflection
coefficients $t_{L,R}$, $r_{L,R}$ as functions of the input wave vector $k$
are obtained from the steady-state solution; the scattering matrix is
unitary $SS^{\dagger }=I_{2}$.

The four-site scattering center is pseudo-Hermitian with the Hamiltonian $%
H_{c}$ and the unitary matrix $q$
\begin{equation}
H_{c}=\omega _{c}I_{4}-\left(
\begin{array}{cccc}
0 & J_{1} & J_{1} & \kappa _{2} \\
J_{1} & -i\gamma & \kappa _{1} & J_{2} \\
J_{1} & \kappa _{1} & i\gamma & J_{2} \\
\kappa _{2} & J_{2} & J_{2} & 0%
\end{array}%
\right) ,q=\left(
\begin{array}{cccc}
1 & 0 & 0 & 0 \\
0 & 0 & 1 & 0 \\
0 & 1 & 0 & 0 \\
0 & 0 & 0 & 1%
\end{array}%
\right) ,
\end{equation}%
where $+i\gamma $ and $-i\gamma $ are the gain and loss; all the couplings $%
J_{1}$, $J_{2}$, $\kappa _{1}$, $\kappa _{2}$ are real numbers. In the
subspaces of connection sites and bulk sites, $q$ satisfies Eq.~(\ref{q}).
Thus, the scattering matrix is unitary and symmetric. Notably, the
scattering is still unitary if the real couplings $J_{1}$ and $J_{2}$ are
simultaneously imaginary.

The transmission and reflection for the unitary scattering in Fig.~\ref{fig2}%
(a) are depicted in Fig.~\ref{fig2}(c). The energy conservation holds after
scattering and the total transmitted and reflected wave probability is
unity; however, the dynamics is non-unitary and the energy conservation is
invalid in the scattering process as numerically simulated in Fig.~\ref{fig2}%
(e) using a Gaussian profile initial excitation $\Omega
^{-1/2}\sum_{j}e^{-(j-n_{0})^{2}\alpha ^{2}/2}e^{-ikj}\left\vert
j\right\rangle $ of the momentum $k$ centered at the site $n_{0}$, where $%
\alpha =0.1$ controls the width, $\Omega $ is the normalization factor, and $%
\left\vert j\right\rangle $ is the basis of port site $j$. The
scattering process begins when the wave packet reaching the scattering
center and the total probability of the excitation starts to change as time
because of the influence of non-Hermiticity. The velocity of the Gaussian
wave packet obtained from the dispersion relation is $dE/dk=2J\sin k$.
The scattering process ends when the wave
packet leaving the scattering center and the total probability no longer
changes as time. The dynamics after scattering reflects the steady-state
solution. The reflected backward going wave packet indicates the reflection
and the transmitted forward going wave packet indicates the transmission~%
\cite{Kim06}.

Alternatively, if each connection resonator is unequally coupled to the gain
and loss resonators as shown in Fig.~\ref{fig2}(b), the four-site scattering
center is still pseudo-Hermitian with $H_{c}$ and $q$%
\begin{equation}
H_{c}=\omega _{c}I_{4}-\left(
\begin{array}{cccc}
0 & J_{1} & J_{2} & \kappa _{2} \\
J_{1} & -i\gamma & \kappa _{1} & J_{2} \\
J_{2} & \kappa _{1} & i\gamma & J_{1} \\
\kappa _{2} & J_{2} & J_{1} & 0%
\end{array}%
\right) ,q=\sigma _{x}\otimes \sigma _{x},
\end{equation}%
where $\sigma _{x}$ is the Pauli matrix. Notably, the unitary transformation
$q$ is not block diagonalized in the subspaces of connection sites and bulk
sites. In this situation, the scattering matrix is non-unitary although the
non-Hermitian scattering center is pseudo-Hermitian. The transmission and
reflection for the non-unitary scattering in Fig.~\ref{fig2}(b) are depicted
in Fig.~\ref{fig2}(d). The scattering is non-unitary in the entire
scattering process and the total probability after scattering is non-unity
as demonstrated in Fig.~\ref{fig2}(f).

The unitary scattering is still possible if the couplings between the ports
and the scattering center are properly redesigned. If both two ports in Fig.~%
\ref{fig2}(b) are simultaneously coupled to sites $1$ and $4$ through the
first sites of the ports at the same strength $-J$, the scattering becomes
unitary. The first sites of two ports effectively become the connection
sites of a six-site scattering center $H_{c}^{\prime }$ and $q^{\prime }$
satisfies Eq. (\ref{q}) with $q_{b}^{\prime }=\sigma _{x}\otimes \sigma _{x}$%
.

\begin{figure}[tb]
\includegraphics[ bb=0 0 400 455, width=8.8 cm, clip]{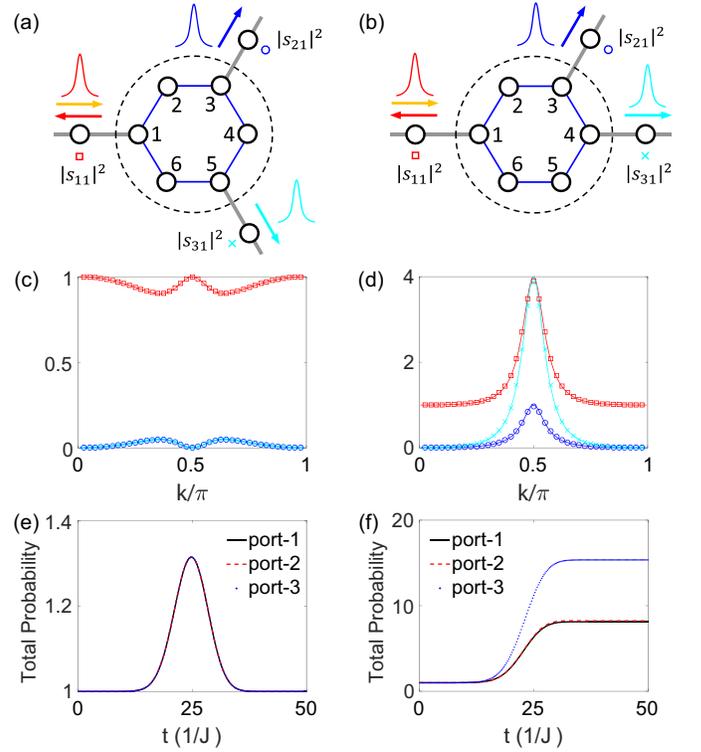}
\caption{ Schematics of the three-port pseudo-Hermitian systems with (a) unitary and (b)
non-unitary scattering. The reflection $|s_{11}|^2$ and transmissions $|s_{21}|^2$, $|s_{31}|^2$ are plotted in (c) and (d) as indicated by the markers in the ports; the curves are analytical results; the markers are numerical simulations. The corresponding total probability in the numerical simulations for $k=\pi /2$ are depicted in (e) and (f). The parameters are $\gamma_j=J$ and $\Delta_j=0$ for all the~sites.}
\label{fig3}
\end{figure}

The unitary scattering is also subtle to the port configuration. We consider
a multi-port pseudo-Hermitian scattering center with six sites and three
ports. The Hamiltonian of the scattering center reads
\begin{equation}
H_{c}=\sum_{j=1}^{6}(\omega _{c}+\Delta _{j})\hat{c}_{j}^{\dagger }\hat{c}%
_{j}+i\gamma _{j}(\hat{c}_{j}^{\dagger }\hat{c}_{j+1}+\hat{c}_{j+1}^{\dagger
}\hat{c}_{j}),
\end{equation}%
where $\hat{c}_{j}$ ($\hat{c}_{j}^{\dagger }$) is the annihilation
(creation) operator for the mode of resonator $j$ and satisfies $\hat{c}%
_{j+6}=\hat{c}_{j}$ ($\hat{c}_{j+6}^{\dagger }=\hat{c}_{j}^{\dagger }$). The
resonator detuning $\Delta _{j}$ is real. The coupling $i\gamma _{j}$ is
non-Hermitian and can be realized through the gain \cite{Fleury18,HCWu20}.
The scattering center $H_{c}$ is pseudo-Hermitian for the unitary matrix $%
q=I_{3}\otimes \sigma _{z}$. In the couple mode theory, the equation of
motion for the light field $\psi _{c,j}$ of the scattering center resonator $%
j$ is%
\begin{equation}
i\dot{\psi}_{c,j}=\left( \omega _{c}+\Delta _{j}\right) \psi _{c,j}+i\gamma
_{j-1}\psi _{c,j-1}+i\gamma _{j}\psi _{c,j+1}.
\end{equation}

If the three ports $1,2,3$ are respectively connected to the odd-site $%
\left\vert 1\right\rangle _{c},\left\vert 3\right\rangle _{c},\left\vert
5\right\rangle _{c}$ (or the even-site $\left\vert 2\right\rangle
_{c},\left\vert 4\right\rangle _{c},\left\vert 6\right\rangle _{c}$) in Fig.~%
\ref{fig3}(a), the additional term $-J\psi _{p,1},-J\psi _{p,2},-J\psi
_{p,3} $ presents in the right side of the equations of motion for the sites
$j=1,3,5$ (or $j=2,4,6$). This pseudo-Hermiticity ensures a unitary
scattering $SS^{\dagger }=I_{3}$. The scattering coefficients are obtained
from the steady-state solution. The transmission and reflection in the three
ports are depicted in Fig.~\ref{fig3}(c). The unity total probability after
scattering reflects the energy conservation as demonstrated in Fig.~\ref%
{fig3}(e), where the initial excitation is that used in the two-port
scattering center. Notably, additional Hermitian couplings presented among
three connections sites will not affect the unitary scattering.

If the three ports are simultaneously connected to the odd-site and
even-site of the scattering center; for example, the three ports are
connected to the sites $\left\vert 1\right\rangle _{c},\left\vert
3\right\rangle _{c},\left\vert 4\right\rangle _{c}$ in Fig.~\ref{fig3}(b).
The scattering is non-unitary. The transmission and reflection in the three
ports are depicted in Fig.~\ref{fig3}(d), the non-unity total probability
indicates the non-unitary scattering and the absence of energy conservation
as observed in Fig.~\ref{fig3}(f). The scattering coefficients diverge and
lasing occurs at the spectral singularity $2\gamma _{j}^{2}=J^{2}$ and $%
\Delta _{j}=0$ for all the six sites \cite{Ali09,LJinPRL}.

All the three constrains in Eq.~(\ref{Constrain}) are satisfied for the
unitary scattering, but at least one of the three constrains is not
satisfied for the non-unitary scattering in the exemplified pseudo-Hermitian
scattering centers. Nevertheless, any pseudo-Hermitian scattering center can
exhibit a unitary scattering if the couplings between the scattering center
and the ports are properly redesigned according to its structure
information, which is completely encoded in the unitary operator $q $. In
this situation, the scattering center is effectively enlarged to include the
first sites of the ports. The original pseudo-Hermitian scattering center $%
H_{c}$ plays the role as the subsystem $H_{b}^{\prime }$ of the enlarged
scattering center $H_{c}^{\prime }$. The couplings between the scattering
center and the first sites of the ports should be reconstructed to satisfy $%
q_{b}^{\prime }H_{eb}^{\prime \ast }=H_{be}^{\prime T}$, where the unitary
operator $q_{b}^{\prime }=q$ defines the pseudo-Hermiticity of the original
scattering center $H_{c}$ with $qH_{c}^{\ast }q^{-1}=H_{c}^{T}$. The unitary
operator $q^{\prime }$ defines the pseudo-Hermiticity of the enlarged
scattering center $H_{c}^{\prime }$
\begin{equation}
q^{\prime }=\left(
\begin{array}{cc}
I_{L} & 0 \\
0 & q%
\end{array}%
\right) ,H_{c}^{\prime }=\left(
\begin{array}{cc}
\omega _{c}I_{L} & H_{eb}^{\prime } \\
H_{be}^{\prime } & H_{c}%
\end{array}%
\right) ,q^{\prime }(H_{c}^{\prime })^{\ast }q^{\prime -1}=(H_{c}^{\prime
})^{T}.
\end{equation}%
After the reconstruction, the effective scattering center $H_{c}^{\prime }$
has the unitary scattering $S^{\prime }S^{\prime \dagger }=I_{L}$ and
ensures the energy conservation.

In addition, the operations of Hermitian conjugation ($\dagger $), complex
conjugation ($\ast $), transpose ($T$), and the unit element constitute a $%
V_{4}$ (also called $D_{2}$) Abelian group. Three operations define the
pseudo-Hermiticity $qH_{c}^{\dagger }q^{-1}=H_{c}$, the time-reversal
symmetry $qH_{c}^{\ast }q^{-1}=H_{c}$, and the reciprocity $%
qH_{c}^{T}q^{-1}=H_{c}$ for the non-Hermitian systems, respectively. The
pseudo-Hermiticity ensures $SS^{\dagger }=I_{L}$, the time-reversal symmetry
ensures $SS^{\ast }=I_{L}$, and the reciprocity ensures $S=S^{T}$ if the
unitary operator $q$ satisfies Eq.~(\ref{q}).

In conclusion, the unitary scattering and the energy conservation are
ensured by certain pseudo-Hermiticity, where the scattering center structure
plays an important role. The pseudo-Hermiticity protected unitary scattering
is independent of the degree of non-Hermiticity. The energy conservation
holds after scattering although it is invalid in the scattering process
under the influence of the non-Hermiticity. We also demonstrate how to
create a unitary scattering in the pseudo-Hermitian system with a
non-unitary scattering through reconstructing the connection couplings. We
unveil the physics of unitary scattering, present novel understanding of
pseudo-Hermiticity, and provide fundamental insight on the energy
conservation in non-Hermitian physics. Our findings are also important
guiding principles promising for the non-unitary scattering. These findings
shed light on the fundamental research and potential applications of
non-Hermitian scattering including the light propagation, mesoscopic
electron transport, and quantum interference \cite{Jiang21}.

\acknowledgments We acknowledge the support of National Natural Science
Foundation of China (Grant No.~11975128).


\begin{thebibliography}{999}
\bibitem{STProof} For a rigorous proof, please see Eq.~(12) in Ref.~\cite{LJinCPL}
and compare the scattering coefficients of the scattering centers $H$ and $H^{T}$. Notice that
the symbols $t_L$, $t_R$, $r_L$, and $r_R$ in Ref.~\cite{LJinCPL} are $s_{nm}$,
$s_{mn}$, $s_{mm}$, and $s_{nn}$ of the scattering matrix with our current notations for any pair of ports $m$ and $n$.
The fact $(A^T)^{-1}=(A^{-1})^T$ is also used in the proof for any square matrix $A$.
\bibitem{SSI} From the off-diagonal term of $SS^{\dagger }$, we obtain $%
r_{L}=-t_{R}r_{R}^{\ast }/t_{L}^{\ast }$. From the diagonal terms of $%
SS^{\dagger }$, we obtain $1=r_{L}r_{L}^{\ast }+t_{R}t_{R}^{\ast }=\left(
r_{R}^{\ast }r_{R}/t_{L}^{\ast }t_{L}+1\right) t_{R}t_{R}^{\ast
}=t_{R}t_{R}^{\ast }/t_{L}^{\ast }t_{L}$.
\end{thebibliography}
\end{document}